# Effect of Intellectual Property Policy on the Speed of Technological Advancement

Ivan D. Breslavsky

*Department of Mechanical Engineering, McGill University,*

*817 Sherbrooke Street West, Montréal, Québec, Canada H3A 0C3*

*Email: ivan.breslavskyi@mcgill.ca*

**Abstract**

In this paper, the agent-based modelling is employed to model the effect of intellectual property policy at the speed of technological advancement. Every agent has inborn preferences towards investing their capital into independent technological development, innovation appropriation, and production. The relative cost of appropriation compared to independent development is chosen as a measure of strictness of intellectual property protection. We vary this parameter and look at the performance of agents with different preferences and overall technological progress. In general, it is found that in the specific setting considered with stronger intellectual property protection leads to faster progress.

# 1. Introduction

Innovation activities have long been recognized as one of the key elements that affect the economic well-being, and therefore, they have received a large amount of attention from researchers. One important question is how much the protection and prohibition of use without permission of inventions accelerate or slow down technological growth. Despite the fact that the first laws on intellectual property protection appeared hundreds of years ago (for example, 15th-century Venetian law (Moser 2013)), the vast majority of countries did not have any intellectual property protection worth mentioning until the second half of the 19th century. Moreover, the mere need for it was questioned. For instance, Thomas Jefferson wrote in 1813 about British copyright protection measures: «…it may be observed that the nations which refuse monopolies of invention, are as fruitful as England in new and useful devices» (Kurland and Lerner 1987). This observation is confirmed by Moser (Moser 2013) based on the results of the Crystal Palace Exhibition in London in 1851 and the American Centennial Exhibition in Philadelphia in 1876. Such development-accelerating effect is labelled as "collective innovation", and occurs when there is a possibility to freely adopt and modify the devices of other inventors, as described by Allen (Allen 1983) and later by Nuvolari (Nuvolari 2004). In modern literature, the abolition or the weakening of both patents and copyrights are increasingly referred to as something that is not only possible but even desirable (Jones 2014; McGeehan 2015; David and Halbert 2015). The papers (von Hippel and von Krogh 2006; Baldwin and von Hippel 2011; Moser 2013) are devoted to the question under which conditions technological development benefits more from stronger intellectual property protection and under which – from collective innovations.

Among the numerous studies on the factors that affect innovations, the mathematical modelling studies have to be mentioned. Mostly, these works are dedicated to the influence on the innovative process of the spatial distribution of the agents (Morone and Taylor 2010; Dunn and Gallego 2010) and the related, but more theoretical, question – the effect of communication network structure (Cowan and Jonard 2003; Carayol and Roux 2005; Zhong and Ozdemir 2010; Pegoretti et al. 2012; Kamath 2013).

Some studies took a game-theoretic and experimental approach to the problem (Suetens 2005; Engel and Kleine 2015). In both studies, it is found that the automatic sharing of innovations increases the innovation investment, contrary to the predictions of the classic economic theory. It is interesting that the experiment carried out in (Engel and Kleine 2015) showed that the innovation investments are higher than what economic theory predicts in all cases, regardless of the intention of the competitors to appropriate innovations. In (Isaac and Reynolds 1992), it is shown that technological development is encouraged by competition rather than by monopoly.

Works on patent race should be mentioned here. Baker and Mezzetti (2005) studied the strategic reasons why firms actually disclose large amount of potentially patentable technical information. Kwon (2012) and Judd et al. (2012) search for optimal patent rules that are the most beneficial to the society. It is concluded (Kwon 2012) that under the condition of small patent propensity strengthening patent protection can decrease firms' incentive to innovate. Panagopoulos (2009) came to close conclusion. He found that the relationship between patent protection and output growth has inverse U shape. The result of the work of Zeira (2011) is that under the competition firm tend to invest more in easy innovations and insufficiently into more expensive one. It is suggested to adjust the patent policies to encourage the investment into long-term and expensive projects. The literature review on advanced of the patent race studies can be found in (Kwon 2012; Judd et al. 2012).

We name here some works that are close in spirit to present work. Hur (2010) built a probabilistic agent-based model to study the dependence of time to invention (that lead to a patent) on risk tolerance of agents competing for the patent. The possibility of technological appropriation was not included into the model. The author comes to a conclusion that competition can expedite the invention of new technology, however, it also increases the total costs spent on the invention. The papers (Gilbert et al. 2001; Sie 2014) are devoted to agent-based investigation of the innovation coalition formations. The spirit of the work (Gilbert et al. 2001) is the closest to present study, but we use simpler approach and focus on intellectual property policy rather than on collaboration of agents and innovation networks formation.

In the present work, we focus on a different question, namely, the effect of the intellectual property protection policy on the speed of technological development. We take an approach that is different from the above-mentioned studies, and use agent-based modelling (Epstein and Axtell 1996; Gilbert and Troitzsch 2005). We do neither take into account the geographic distribution of agents nor the structure of the communication network, assuming that the technological information of any agent is available to others, but appropriation has policy-dependent costs. Each agent is an economic enterprise that has the disposition to invest certain shares of profit in production, independent innovation development, or appropriation of innovations from other agents. We prescribe certain cost of innovation appropriation (resulting from the intellectual property protection policy), and then look at the performance of the agents with different strategies and the overall technological growth rate. Due to the lack of quantitative data, this study is a pure "complex thought experiment", as described by Cederman (1997), aiming to get a qualitative understanding on the policy implications, rather than focusing on quantitative results.

**2. Problem Statement**

Let us consider a population of competing agents. They produce and sell a product, the total money earned by an agent every turn grows with the quantity of products and with the level of technology used in the production process. Two adjacent types of technologies are considered – technology A and technology B. These technologies are not independent in a sense that the higher the level of technology B possesses the agent, the simpler is the development of the next level of technology A and vice versa. We take into consideration two technologies instead of one to check the following hypothesis: acceleration of technological development could result from the fact that particular technology is developed by the agent who has a higher capability for development of this technology and then in the case of cheap appropriation is "donated" to the public.

Every agent is born with dispositions to spend certain parts of his earnings on production, independent development of technology, and appropriation of technology from others. The latter can be more or less expensive than the independent development, depending on the legislation. The measures that could be taken by technology owners for preventing the spread of technology are not considered here. The effect of cost-of-appropriation to cost-of-innovation ratio on the speed of technical progress is the primary question of the present study.

Each agent has the following characteristics:

$K_i > 0$ – capital (this characteristic changes from turn to turn).

$x_i$ – part of the agent's capital that he spends on innovation, $0 \leq x_i < 1$ (this value is fixed for the agent).

$y_i$ – part of the agent's capital that he spends on production, $0 < y_i \leq 1$ (fixed); $1 - x_i - y_i$ is the rest of capital, which the agent spends on the appropriation of technology, $0 \leq 1 - x_i - y_i < 1$.

$InnA_i$ – agent's level of technology A (changes), $InnA_i \geq 1$.

$InnB_i$ – agent's level of technology B (changes), $InnB_i \geq 1$.

$CapA_i$ – agent's capability for development of technology A (fixed), $0.1 \leq CapA_i \leq 1$. Similarly,

$CapB_i$ – agent's capability for development of technology B (fixed), $0.1 \leq CapB_i \leq 1$.

We use the assumption that the total capital stays constant over all the turns of the simulation. The agent's profit is the share of total capital that is proportional to the quantity of goods produced and to the product $InnA_i * InnB_i$. The latter reflects the idea that more advanced goods cost more than the low-end goods.

The agent, whose capital falls below a certain threshold is eliminated from the population.

Each agent uses a designated part of his capital ($x_i$) for the development of new technology. Every turn the agent decides in which technology development (A or B), he is going to invest. The decision is made based on the increase-of-profit to turns-to-next-level ratio (assuming that the agent's capital and all other factors like the technological levels of the rest of agents will remain the same over the next turn). Price for every next technological level grows quadratically, more specifically, the price of the development of the technology level *numI* is equal to $VI*(numI-1)^2$, where *VI* is a specific cost of innovation, in this study assumed to be 1. This law of innovation price increase is the same for both technologies, A and B.

The assumption of the quadratic growth of the cost of innovation is speculative and should be improved in further studies, but so far, the author was not able to find the estimation of this function in literature.

Another assumption related to technology development includes two quantities that increase the rate of development (they are used as multipliers before the capital investment in the innovation): $CapA_i$ or $CapB_i$ for technology A and B, correspondingly; and $(1+\ln InnA_i)$ $(1+\ln InnB_i)$ for either of the technologies, A or B. The latter multiplier represents the mutual dependence of technologies.

The agent, who spends a part of his capital on technology appropriation ($x_i + y_i < 1$), acts according to the following algorithm: he looks whether there are agents with more advanced technology than his. If not, he spends the part of capital intended for appropriation ($1 - x_i - y_i$) on production. If yes, but only one of the technologies (A or B) is more advanced, he spends money on the appropriation of that technology. If higher levels of both technologies are present in the population, then he adopts the one that is weaker (this is a rational choice in the considered problem formulation). If both adopters' technological levels are the same, he adopts the one to which he has the smallest capability.

The price of appropriation is assumed to be flat for both technologies and all technological levels. We introduce the parameter *Vs as a* measure or relative difference between the price of appropriation and independent development. The total price that the agent has to pay to adopt the technological level *numI* (provided he has a technological level *numI*–1 ) is $(VI/Vs)*(numI–1)^2$. The question of interest in this study is how the *VI /Vs* ratio affects the speed of technological development in the society. Since we take VI=1 in this study, Vs>1 means that it is cheaper to adopt new technology than to develop it. Likewise, Vs<1 means that it is cheaper to develop independently than to adopt it (for example, due to the high possible penalties of the intellectual property violation or to the strong common measures of know-how protection).

All calculations are implemented using Mathematica software (Wolfram Mathematica 2016). The flowcharts of the agent's logic are given in Appendix.

## 3. Results and discussion

In the numerical investigations for the sake of convenience, the agents were divided into four groups, depending on their preferences:

1. Pure innovators: $x_i>0$, $x_i+y_i=1$.

2. Pure adopters: $x_i=0$, $y_i<1$.

3. Innovators-adopters $x_i>0$, $x_i+y_i<1$.

4. Pure producers $x_i=0$, $y_i=1$.

Initial conditions were given to the agents on a random basis. Initial capital was uniformly distributed, the richest agents having 4 times larger capital than the poorest ones. Afterwards, the capital was normalized to a total capital of 100.

The agent continues to operate, until his capital is lower than a certain threshold (taken 0.1 in the present study), after which the agent is considered to be broken-down and is eliminated from the consideration.

All agents start from the same (first) technological level in both technologies.

We considered five cases: from $Vs=0.3$ (corresponding to the highest cost of appropriation compared to independent invention) to $Vs=3$ (corresponding to the lowest cost of appropriation compared to independent invention) with intermediate values of $Vs=0.5$, $Vs=1$ and $Vs=2$.

In every case, we run the model 200 times to obtain the average characteristics, such as the technological level and the most successful strategies, for a given value of each parameter $Vs$. Each run involves 50 time steps, which was found to be enough to finish the competition and to find the winner, in most cases.

In case of $Vs=0.3$ (highest cost of appropriation), the average level of technology is 24 with absolute maximum over all run equally to level 36 and absolute minimum equal to $8^{th}$ level. The mean deviation in technology level in this case is 4.5.

For this value of $Vs$ there is only one surviving agent, who is almost always the inventor (group 1) and spends 10-30% (i.e. with $0.1 \leq x_i <0.3$) of his capital on innovations. In a couple of simulations, the winner was from group 3 (innovators-adopters).

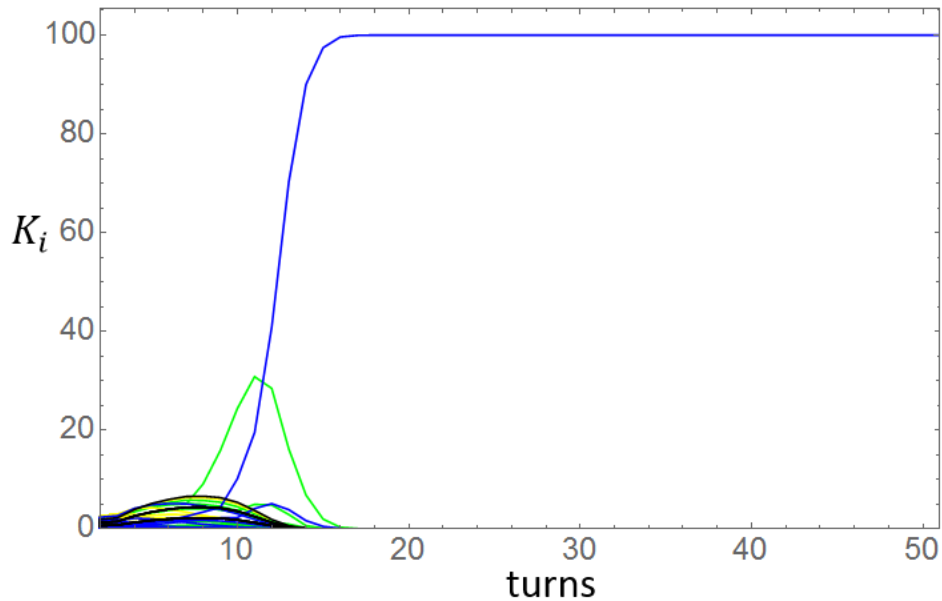

**Fig. 1.** The typical evolution of the system for high relative cost of appropriation ($Vs=0.3$). Pure innovators (group 1) are shown in green, pure adopters (group 2) in yellow, innovators-adopters (group 3) in blue, pure producers (group 4) in black.

The typical for this case plot of capital vs. time is given in Fig. 1. Hereinafter pure innovators (group 1) are shown in green, pure adopters (group 2) in yellow, innovators-adopters (group 3) in blue, pure producers (group 4) in black.

In the case of $Vs=0.5$, the overall picture is very similar to the previous one. The average level of technology is 24. The level varied in the range [9, 36] and with a mean deviation of 4.5.

The next case of $Vs=1$ is quite different from the previous ones. Maximal achieved technological level varied in a range [2, 35] with an average value of 20 and a mean deviation of 6.8.

Typical capital vs. time plot for this case is given in Fig. 2.

In most cases, there is only one surviving agent, but now, it can belong to any of the groups 1-3, i.e. pure innovators, pure adopters, and innovators-adopters.

In the minority of runs, there are a few survivors. These are the situations where the locking of the system in a low-technology regime is observed (see Fig. 3). This happens when all the agents willing to spend money on the innovations are ruined by adopters.

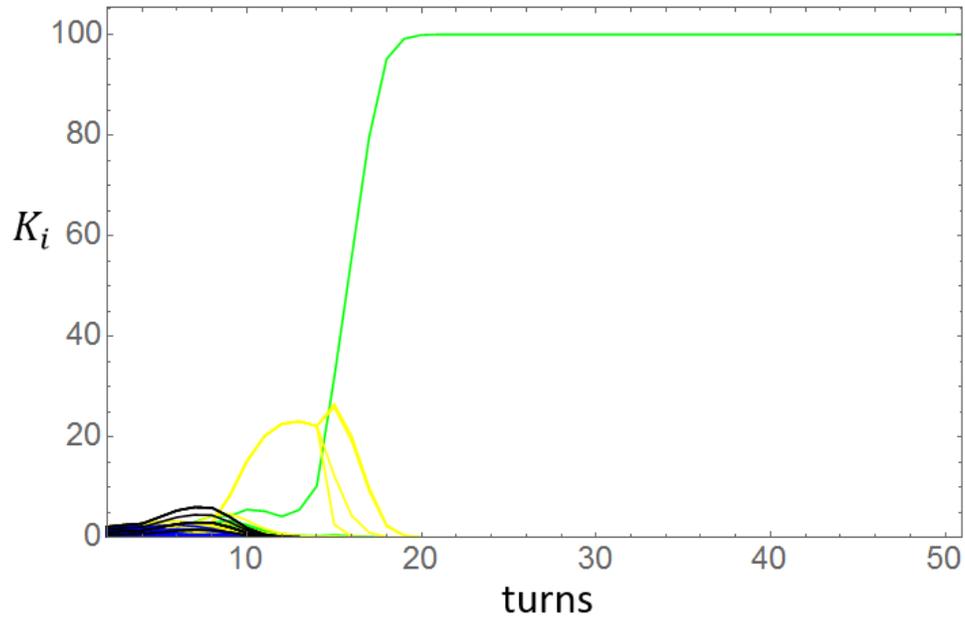

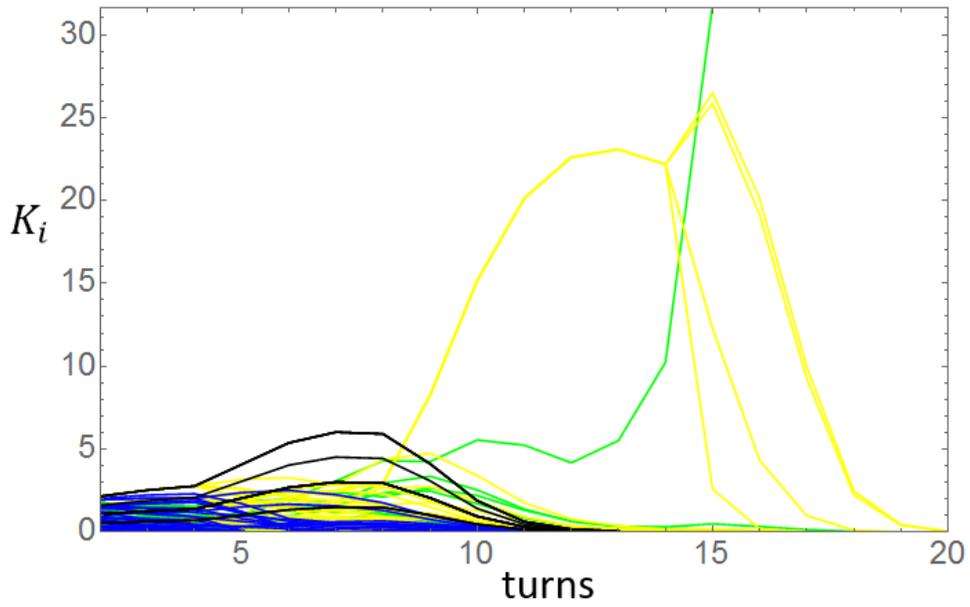

**Fig. 2.** The typical evolution of the system for medium relative cost of appropriation ($Vs=1$). (b) is the magnification of the (a). Pure innovators (group 1) are shown in green, pure adopters (group 2) in yellow, innovators-adopters (group 3) in blue, pure producers (group 4) in black.

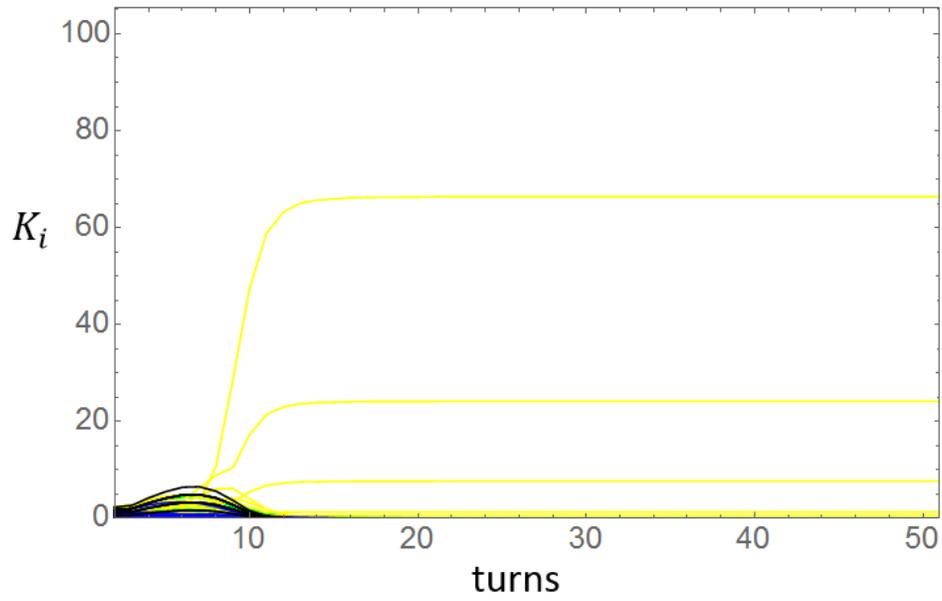

(a)

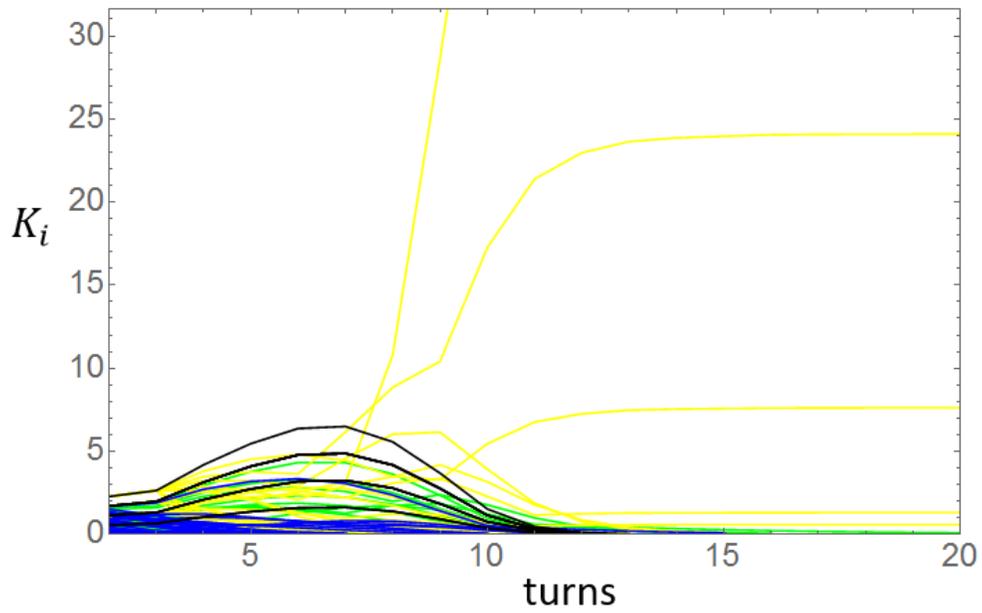

(b)

**Fig. 3.** Rare case of the system locking in the low-tech area, $Vs=1$. Both technologies A and B have level 2 with no possibility to grow. (b) is the magnification of the (a). Pure innovators (group 1) are shown in green, pure adopters (group 2) in yellow, innovators-adopters (group 3) in blue, pure producers (group 4) in black.

The figures for the case of $Vs=2$ is not very different from the case of $Vs=1$. Maximal achieved technological level varied in a range [2,35] with an average value of 15 and a mean deviation of 8.3.

Again, "low-tech locking" (similar as shown on Fig. 3) is present in some cases. It should be noted that, unlike the case without locking, there are multiple survivors in this situation. The steady state is reached, when all the survivors have a stable share of marked.

And the last considered case is $Vs=3$, which is the situation of the lowest cost of appropriation. Maximal achieved technological level varied in a range [2,33] with an average value of 14 and a mean deviation of 8.5. There are two typical outcomes:

1. Single winner case (pure innovator or innovator-adopter spending 2-35% of capital on innovations) that constituted roughly four-fifths of all runs (Fig. 4);

2. The rest of runs returned multiple-survivors result (all adopters) with locking in a low-technological regime. The evolution in time of capital distribution for this case is given in Fig. 5.

It should be noted that pure producers (group 4) do not survive under any values of cost-of-appropriation to cost-of-innovation ratio.

## 4. Conclusions

In this paper, we proposed a model for investigation of the influence of the technology appropriation ease at the speed of technological development. Certain publications claim that lowering the barriers for the technology spread among market players can accelerate the technological progress. However, the opposite effect is actually observed. The more expensive it is to adopt the technology from others, the faster the progress. Our conclusion conforms to the conclusions made in paper (Saam and Kerber 2013) where it was shown that decentralization of the experimentation in longer run leads to a better performance and reduces the risk of low overall performance.

The present study is the first known approach on the topic and thus, the model so far has multiple limitations and involves assumptions that should be verified in the future. One possible way to improve the model is to include the possibility of new agents coming into the market and study of the case, when the direction of the promising innovations is not known (this can be done, for example, by adopting the approach from (Gilbert et al. 2001)). These are the author's goals for the near future.

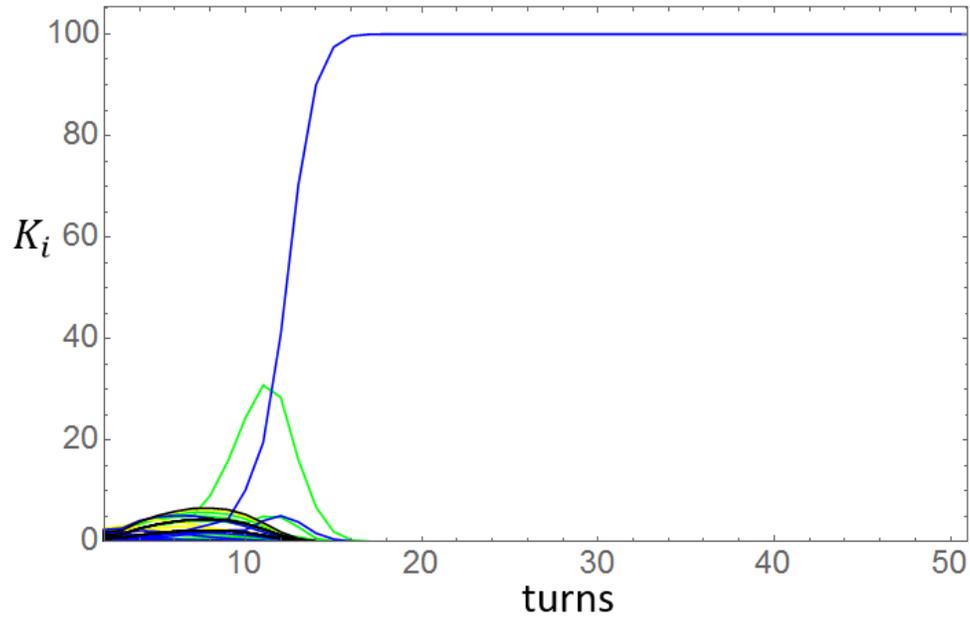

(a)

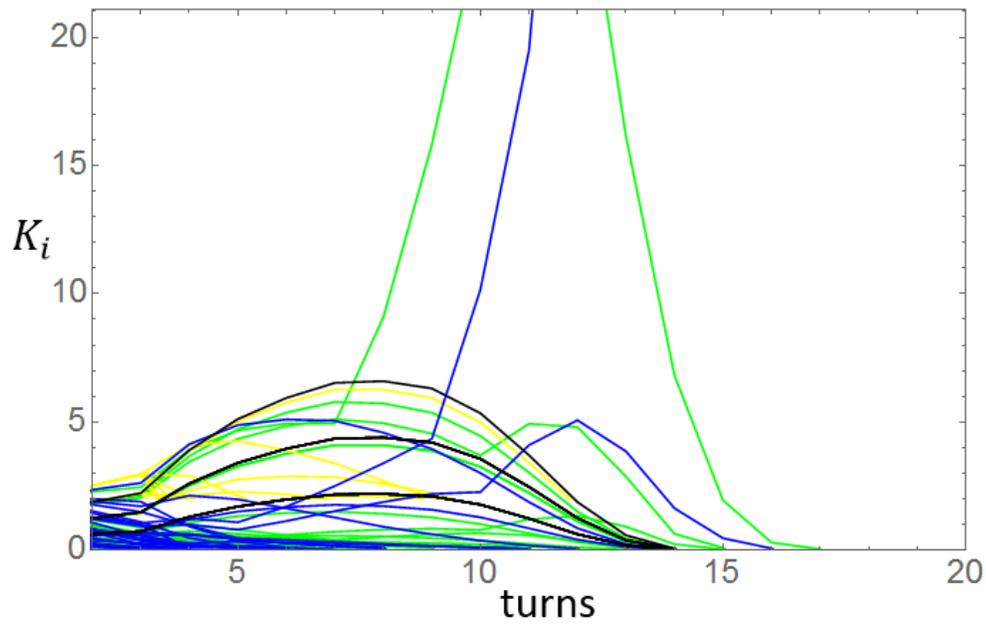

(b)

**Fig. 4.** Single-survivor case of the system for low relative cost of appropriation ($Vs=3$). (b) is the magnification of the (a). Pure innovators (group 1) are shown in green, pure adopters (group 2) in yellow, innovators-adopters (group 3) in blue, pure producers (group 4) in black.

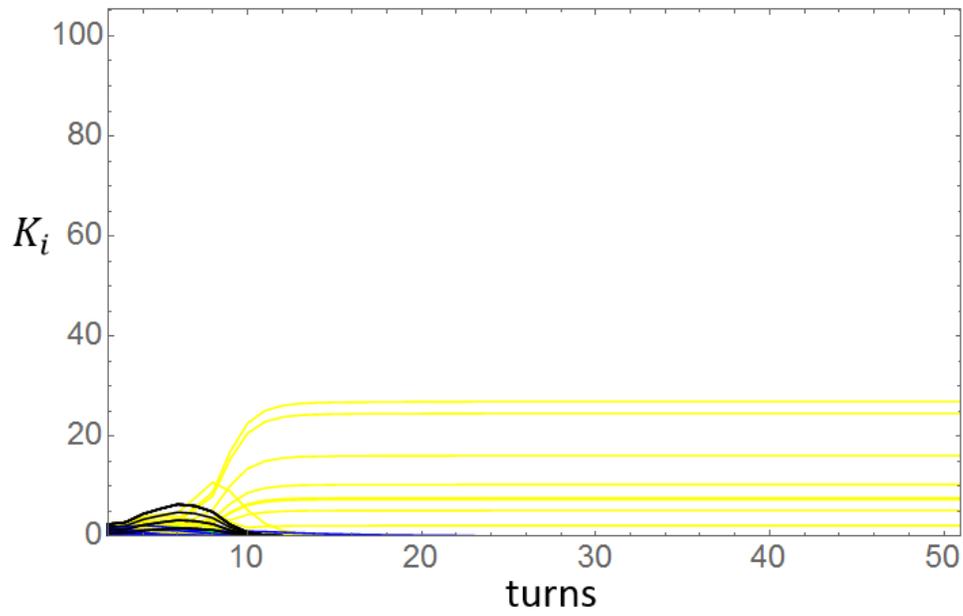

(a)

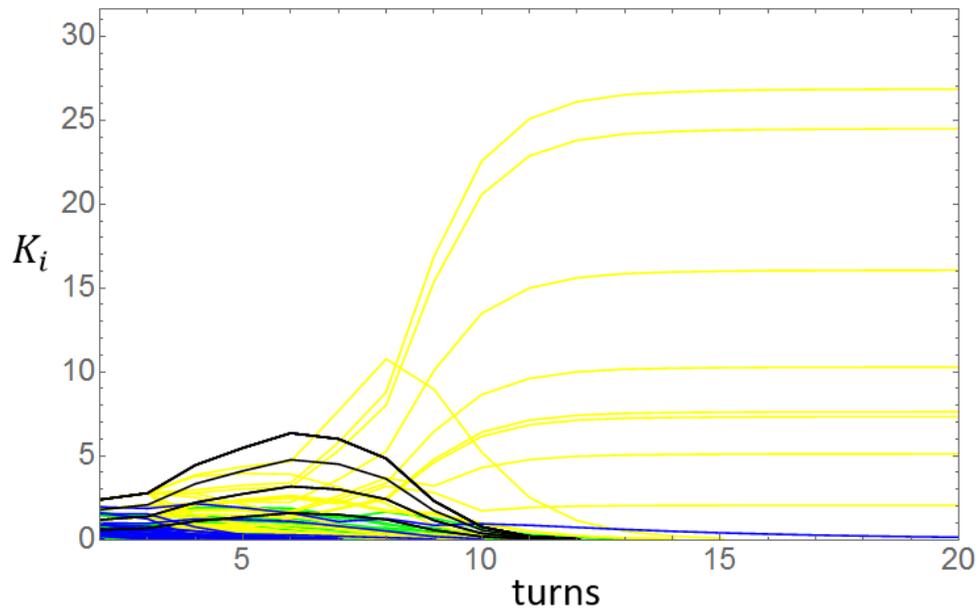

(b)

**Fig. 5.** The typical evolution of the system – locking with technology levels equal to 2 – for low relative cost of appropriation ($Vs=3$). (b) is the magnification of the (a). Pure innovators (group 1) are shown in green, pure adopters (group 2) in yellow, innovators-adopters (group 3) in blue, pure producers (group 4) in black.

# Appendix. Flowcharts

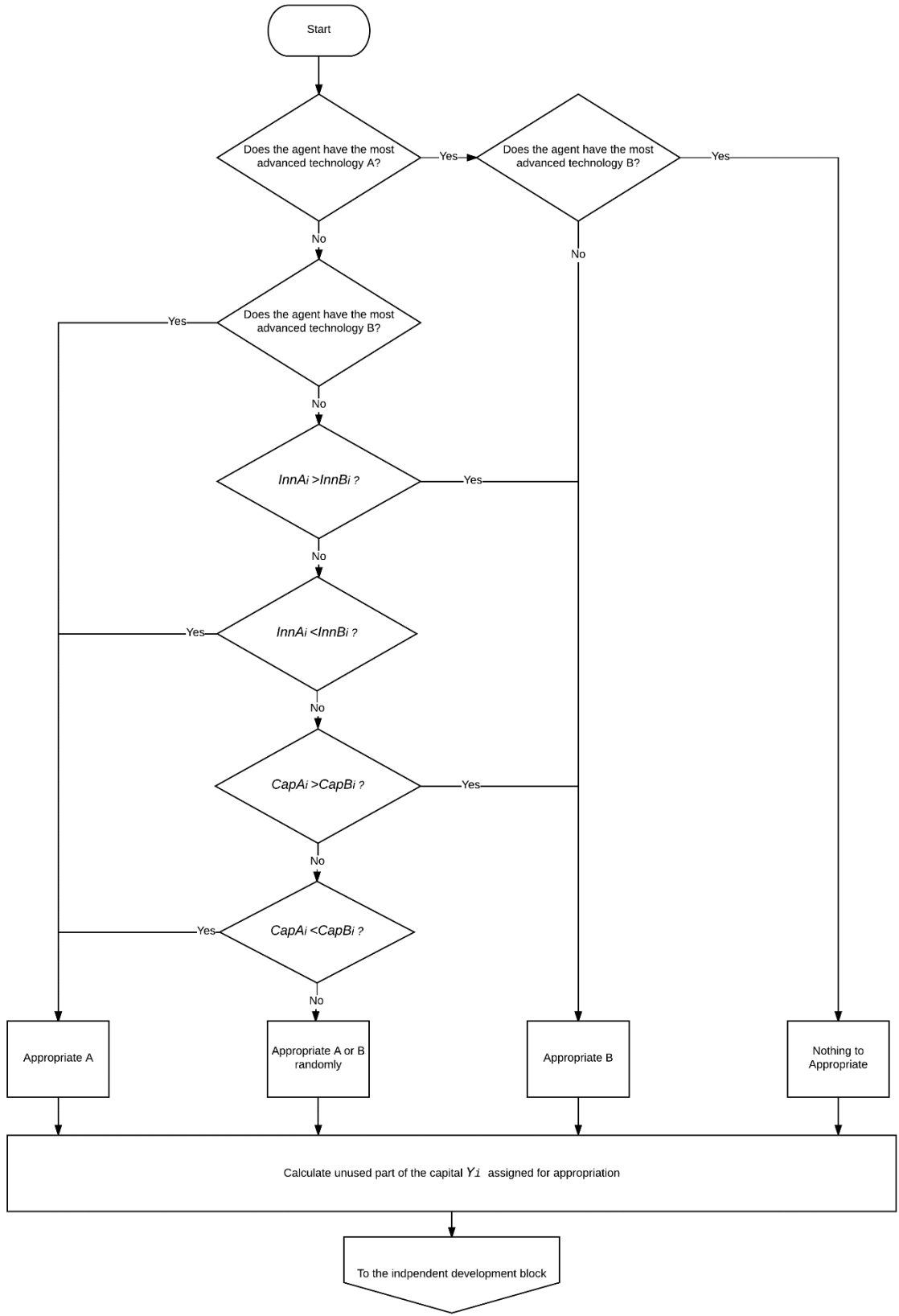

**Fig. A1.** The flowchart of the agent's appropriation block.

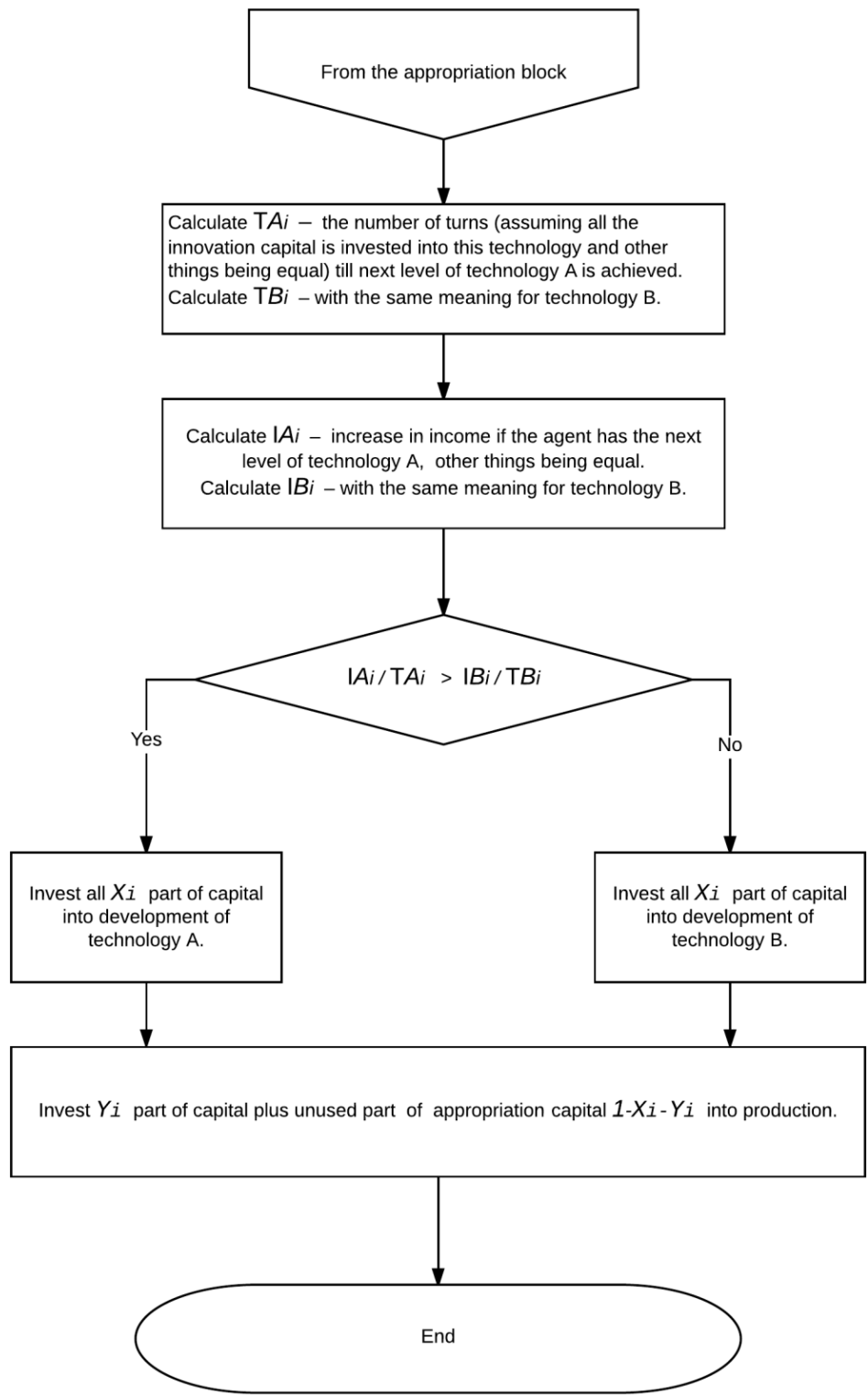

**Fig. A2.** The flowchart of the agent's innovation and production block.